\documentclass[11pt]{article}
\usepackage{amssymb}
\usepackage{amsmath}

\newcommand{\hko}{\hookrightarrow}
\newcommand{\pa}{\partial}
\newcommand{\n}{\nonumber\\}

\newcommand{\clp}{\cl_{p,q}}
\newcommand{\bec}{\begin{center}}
\newcommand{\eec}{\end{center}}

\newcommand{\rf}{\lrcorner}
\newcommand{\bea}{\begin{array}}
\newcommand{\ear}{\end{array}}
\newcommand{\cl}{C\ell}
\newcommand{\om}{\omega}
\newcommand{\DDD}{\mathbb{D}}
\newcommand{\noi}{\noindent}\newcommand{\Ra}{\rightarrow}

\newcommand{\me}{\frac{1}{2}}
\newcommand{\RR}{\mathbb{R}}\newcommand{\op}{\oplus}

\newcommand{\ZZ}{\mathbb{Z}}
\newcommand{\ot}{\otimes}
\newcommand{\ww}{{\bf w}}
\newcommand{\la}{\Lambda}
\newcommand{\bege}{\begin{equation}}
\newcommand{\enge}{\end{equation}}
\newcommand{\w}{\wedge}
\newcommand{\g}{\gamma}
\newcommand{\sa}{\underline{\star}}\newcommand{\saa}{\overline{\star}}

\newcommand{\beq}{\begin{eqnarray}}\newcommand{\benu}{\begin{enumerate}}\newcommand{\enu}{\end{enumerate}}
\newcommand{\eeq}{\end{eqnarray}}
\newcommand{\mt}{\mathcal}

\newcommand{\vv}{{\bf v}}
\newcommand{\ee}{{\bf e}}\newcommand{\aaa}{{\bf a}}\newcommand{\bb}{{\bf b}}
\newcommand{\uu}{{\bf u}}
\newcommand{\CC}{\mathbb{C}}

\newcommand{\mrr}{\mathring}
\newcommand{\ul}{\underline}
\newcommand{\vbn}{\blacktriangleleft}
\newcommand{\vvn}{\blacktriangleright}

\newcommand{\up}{\Upsilon}
\newcommand{\bla}{{\breve{\la}}}
\newcommand{\bx}{\begin{pmatrix}}
\newcommand{\ex}{\end{pmatrix}}
\newcommand{\vcx}{\varepsilon}

\newcommand{\mme}{{\mathfrak{e}}}

\usepackage{indentfirst}  
\begin{document}
\title{Extended Grassmann and Clifford algebras}
\author{{\bf R. da Rocha}\thanks{Instituto de F\'{\i}sica Gleb Wataghin (IFGW), Unicamp, CP 6165, 13083-970, Campinas (SP), Brazil. 
e-mail: {\tt roldao@ifi.unicamp.br}. Supported by CAPES.}
\and
{\bf J. Vaz, Jr.}\thanks{Departamento de Matem\'atica Aplicada, IMECC, Unicamp, CP 6065, 13083-859, Campinas (SP), Brazil. 
 e-mail: 
{\tt vaz@ime.unicamp.br}.}}

\date{}\maketitle

\abstract${}^{}$\begin{center}
\begin{minipage}{12cm}$ \;\;\;\;\;$
This paper is intended to investigate Grassmann and Clifford algebras over Peano spaces, introducing their 
respective associated extended algebras, and to explore these concepts also from the counterspace viewpoint. 
 The presented formalism explains how the concept
 of chirality stems from the bracket, as defined by Rota et all \cite{rota}. The exterior (regressive) algebra
is shown to share the exterior (progressive) algebra in the direct sum of chiral and achiral subspaces.
The duality between scalars and volume elements, respectively under the progressive and
the regressive products is shown to have chirality, in the case when the dimension $n$ of the Peano space is even.
In other words, the counterspace volume element 
is shown to be a scalar or a pseudoscalar, depending on the dimension of the vector space to be respectively odd or even.
The de Rham cochain associated with the differential operator 
 is constituted by a sequence of exterior algebra homogeneous subspaces subsequently chiral and achiral. 
Thus we prove that the exterior algebra over the space and the exterior algebra constructed on
the counterspace are only \emph{pseudoduals} each other, if we introduce chirality.
 The extended Clifford algebra is introduced 
in the light of the periodicity theorem of Clifford algebras context, wherein the Clifford and extended Clifford algebras $\clp$
can be embedded in $\cl_{p+1,q+1}$, which is shown to be exactly the extended Clifford algebra. 
We present the essential character of the Rota's bracket, relating it to the formalism exposed by 
Conradt \cite{con},  introducing the regressive product and subsequently  the counterspace. 
Clifford algebras are constructed over the counterspace, and the duality between  progressive and 
regressive products is presented using the dual Hodge star operator. The differential and codifferential
operators are also defined for the extended exterior algebras from the regressive product viewpoint, and it is shown they 
uniquely tumble right out  
progressive and regressive exterior products of 1-forms.

\end{minipage}\end{center}
\bigskip\medbreak\noi
Key words: Grassmann and Clifford algebras, chirality, duality, counterspace.
\medbreak\noi  MSC classification: 15A66.

\section*{Introduction} 
{Grassmann\footnote{Grassmann algebras are defined as exterior algebras endowed with a metric structure.
In this sense exterior algebras are vector spaces (endowed with the wedge product) devoid of a metric.}
 and Clifford algebras have played an essential role in modern physics (see, e.g., \cite{bay, bt, hes})  since
 their discovery \cite{gra}. This work is intended to give a precise mathematical formulation of the 
concept of chirality associated with these algebras, which is defined as a multiplication by a pseudoscalar\footnote{Pseudoscalars are scalars that change sign under orientation change, and this denomination is not to be confused with pseudoscalars as elements of $\Lambda^n(V)$.} $\vcx$ 
 that satisfies $\vcx^2 = 1$.  The mathematical 
 description of chirality has fundamental importance, particularly in the context of the extended Clifford algebras to be presented here, and it may bring a deep 
understanding of Nature. Our viewpoint shed some new light on these old fundamental concepts, and can be applied immediately to physics.
 In particular the formulation of electromagnetic theory  \cite{bay, misn} in this formalism is 
more natural, correct, precise and geometrically sensible if  differential forms intrinsically endowed with chirality  are used.
The metric-free formulation of electrodynamics brings a geometric character and a clear physical interpretation, and the formalism exhibited in some 
manuscripts \cite{janc} motivates the formulation using the Rota's bracket.
The (Rota's) bracket, a pseudoscalar that gives chirality to differential forms and multivectors in the Grassmann and Clifford algebras over a Peano space, 
is presented.
Although the extended Grassmann algebra formalism has a didactic explanation for instance in  Jancewicz's paper \cite{janc}, 
 in the light of the Rota's bracket this
formalism can be alternatively explored, by introducing the regressive product \cite{gra, con}. 
 After defining chiral
 differential forms, which changes sign under orientation change, the extended exterior algebra is defined and discussed,
 constructed as a direct sum of two copies (chiral and achiral) of exterior algebras. We present the quasi-Hodge dual star operators
and their chiral partners. After introducing a metric in the Peano space, the extended Grassmann and Clifford  algebras are introduced together with 
the chiral dual Hodge star operators. The regressive product
 is defined together with the concept of counterspace, preserving the 
principle of duality \cite{con, hest3}. An analogue of the Morgan law to the Grassmann-Cayley algebra, defined
 to be the Grassmann extended algebra endowed with
the regressive product, is also investigated, and the counterspace volume element is shown to be scalar {\it or} pseudoscalar, depending on the space dimension to be, respectively, odd {\it or}
even. The de Rham cochain, generated by the codifferential operator related to the regressive product,  is composed  
by a  sequence of exterior algebra homogeneous subspaces that are subsequently chiral and achiral. This is an astonishing  
character of the formalism to be presented, since the duality between exterior algebras associated respectively with  
the space and counterspace is irregular,
in the sense that if we take the exterior algebra duality associated with the space, we obtain the exterior algebra associated with the counterspace, but 
the converse produces the space exterior algebra, which homogeneous even [odd] subspaces  are chiral [achiral], depending on the original vector space
dimension (see eq.(\ref{ritaguedes}) below). 
The present formalism explains how the concept of chirality stems from the bracket defined by G.-C. Rota \cite{rota}. 

Denoting $\mrr{V}$ the chiral vector space associated with $V$, the embedding of a vector space $V$ in the vector space $V\op \mrr{V}$ is necessary in order
 to be possible to correctly  
introduce  extended Clifford algebras. The units respectively associated with the field $\RR$ and $\mrr\RR$, over which $V$ and $\mrr{V}$ are constructed, 
are considered to be distinct, as the metrics in each one of these spaces. 
Besides, the metric in $V\op \mrr{V}$ that takes values in distinct subspaces of $V\op\mrr{V}$ is defined to be identically null, 
otherwise it can be shown several inconsistencies in the formulation. 
In a natural manner, the metric in $V\op\mrr{V}$ is the (direct) sum of the metrics in  $V$ and $\mrr{V}$. The unit 
of $V\op\mrr{V}$ is the sum of the units of $V$ and $\mrr{V}$. 
As $V\simeq\RR^{p,q}$, each of the  objects acting on $V\op\mrr{V}$ 
are shown to be elements of the Clifford algebra $\cl_{p+1,q+1}$, which is essentially the extended Clifford algebra.

The regressive product is introduced together with the counterspace, providing a formal pre-requisite  to 
define  Clifford algebras over the counterspace \cite{con}, illustrating in this way the counterspace dual character. 
Dualities and codualities are defined in space and counterspace, from the use of the dual Hodge star operator.  
The dual character of contraction operators, defined in space and counterspace, is also established, following Conradt's route \cite{con}. 
The codifferential operator is uniquely defined in terms of 
the regressive and progressive exterior products.

 This paper is organized as follows: after presentiing algebraic preliminaries in Section 1, in Section 2 the Rota's bracket and
Peano spaces are introduced. In Section 3 we present the extended exterior algebra and the chiral
 quasi-Hodge star operators. In Section 4 the 
extended Grassmann and Clifford algebras, in the context of the Periodicity Theorem, are defined, and in Section 5 the embedding of a vector space in the 
extended vector space is considered, after which in Section 6 the regressive product is introduced.
 In Section 7 the chiral counterspace is defined and investigated together with its constituents, the differential coforms. Besides, the counterspace volume
element with respect to the regressive product is undefined to be a scalar or a pseudoscalar until we specify whether the dimension of the Peano space
is respectively odd or even, by showing that  the  volume element constructed from a cobasis of the counterspace is a scalar or a pseudoscalar, 
depending on the vector (Peano) space dimension. In Section 8 Clifford algebras over the counterspace are constructed, and in Section 9 the duality and coduality principles 
are introduced in space and counterspace, showing a close relation involving  the regressive and progressive products, and the dual Hodge star 
operator. Also, the contraction in counterspace, after defined, is investigated in the light of the duality and coduality. Finally in Section  10
the differential and codifferential  operators are introduced in the counterspace context, in an alternative extended formalism. 

\section{Preliminaries}
Let $V$ be a finite $n$-dimensional real vector space. We consider the tensor algebra $\bigoplus_{i=0}^\infty T^i(V)$ from which we
restrict our attention to the space $\Lambda(V) = \bigoplus_{k=0}^n\Lambda^k(V)$ of multicovectors over $V$. $\Lambda^k(V)$
denotes the space of the antisymmetric
 $k$-tensors, the $k$-forms.  Given $\psi\in\Lambda(V)$, $\tilde\psi$ denotes the \emph{reversion}, 
 an algebra antiautomorphism
 given by $\tilde{\psi} = (-1)^{[k/2]}\psi$ ([$k$] denotes the integer part of $k$). $\hat\psi$ denotes 
the \emph{main automorphism or graded involution},  given by 
$\hat{\psi} = (-1)^k \psi$. The \emph{conjugation} is defined as the reversion followed by the main automorphism.
  If $V$ is endowed with a non-degenerate, symmetric, bilinear map $g: V\times V \rightarrow \RR$, it is 
possible to extend $g$ to $\la(V)$. Given $\psi=\uu_1\w\cdots\w \uu_k$ and $\phi=\vv_1\w\cdots\w \vv_l$, $\uu_i, \vv_j\in V$, one defines $g(\psi,\phi)
 = \det(g(\uu_i,\vv_j))$ if $k=l$ and $g(\psi,\phi)=0$ if $k\neq l$. Finally, the projection of a multivector $\psi= \psi_0 + \psi_1 + \cdots + \psi_n$,
 $\psi_k \in \la^k(V)$, on its $p$-vector part is given by $\langle\psi\rangle_p$ = $\psi_p$. 
The Clifford product between $\ww\in V$ and $\psi\in\la(V)$ is given by $\ww\psi = \ww\w \psi + \ww\cdot \psi$.
 The Grassmann algebra $(\la(V),g)$ 
endowed with this product is denoted by $\cl(V,g)$ or $\cl_{p,q}$, the Clifford algebra associated with $V\simeq \RR^{p,q},\; p + q = n$.

\section{Peano  spaces}

 Let  $V$ be a $n$-dimensional vector space over a field\footnote{Here we shall consider any field $\mathbb{F}$ with characteristic
different of two. In particular the complex field $\CC$ could be used, but we prefer to use $\RR$ in order to clarify the concepts to
be introduced.} $\RR$. A basis $\{\ee_i\}$ of  $V$ is chosen, and
   $V^*$  denotes the dual space associated with $V$, which has a basis 
$\{{\bf e}^i\}$ satisfying $ \ee^i(\ee_j) = \delta^i_j.$ Since  ${\rm dim} \;V^* = {\rm dim} \;V$, there exists a non-canonical isomorphism between  $V$ and $V^*$.
A Peano space is a pair ($V, [\;\;]$), where [\;\;] is  an  alternate $n$-linear form over $\RR$, the {\it bracket}, defined as the map
$[\;\;]: \underset{n\;\text{times}}{\underbrace{V\times V\times \cdots\times V}} \rightarrow \RR$  with the properties:
\begin{enumerate}\label{enuu}
\item For all $\ww_1, \ww_2 \in V$ and $\mu, \nu \in \RR$,\\
$[\vv_1, \ldots, \vv_{i-1}, \mu\ww_1 + \nu\ww_2, \vv_{i+1}, \ldots, \vv_n] = 
\mu[\vv_1, \ldots, \vv_{i-1}, \ww_1, \vv_{i+1}, \ldots, \vv_n]$\\ $\;{}\hspace*{7cm}+ \nu[\vv_1, \ldots, \vv_{i-1}, \ww_2, \vv_{i+1}, 
\ldots, \vv_n]$;
\item $[\vv_1, \vv_2, \ldots, \vv_n] = {\rm sign} (\sigma) [\vv_{\sigma(1)}, \vv_{\sigma(2)}, \ldots, \vv_{\sigma(n)}],$ where $\sigma$ is
a permutation of the set $\{1,2,\ldots,n\}$.
\end{enumerate}
Indeed, the bracket is an element of $\Lambda^n(V)$. 
A Peano space is called \emph{standard} if there exists a basis $\{\uu_i\}$ of vectors in $V$ such that  $[\uu_1, \uu_2, \ldots, 
\uu_n] \neq 0$ \cite{fauser}. Unless otherwise stated we assume standard Peano spaces and they shall be denoted uniquely by $V$. 
The vectors  $\ww_1$ and $\ww_2$ are  linearly independent if there exists $n-2$ vectors $\uu_3,\ldots, \uu_n$ such that $[\ww_1, \ww_2, \uu_3, \ldots, \uu_n] \neq 0.$ 
 If another basis $\{\vv_i\}$ of $V$ is taken, its bracket in terms of the bracket computed at the basis $\{\ee_i\}$ is given by
$[\vv_1, \vv_2, \ldots, \vv_n] = {\rm det}(v_i^j)\;[\ee_1, \ee_2, \ldots, \ee_n]$, where $\vv_i = v_i^j\ee_j$.  The number 
${\rm det}(v_i^j)$ is positive [negative] 
if the bases $\{\ee_i\}$ and $\{\vv_i\}$  have the same [opposite] orientation, where the orientation 
in $V$ is defined as a choice in $\ZZ_2$ of an equivalence class of a basis in $V$, i.e., two basis are equivalent if they have the same orientation. 
Any space $V$ has only two possible orientations, according to the sign of ${\rm det}(v_i^j)$ and any even permutation  
of  basis elements induces the same orientation. A basis $\{\ee_i\}$ can be transformed as $\ee_i\mapsto A\ee_i$, where  $A\in {\rm Hom}(V)$ is a preserving orientation 
homomorphism, 
and it still represents the same orientation, since the bracket is non-null.
  The basis $\{\ee_i\}$ is denominated a unimodular basis if $[\ee_1, \ee_2, \ldots, \ee_n] = 1$. But as Rota pointed out \cite{rota}, 
the orientation can be equivalently defined by an ordered sequence of vectors entering the bracket, and for instance 
$[\ee_2, \ee_1, \ldots, \ee_n] = -1$ (following from the 2$^{\rm nd}$ property of bracket definition above). 
Then the value assumed by the bracket on an ordered sequence of [unit] vectors can assume 
negative and positive values [$\pm 1$], defining two equivalence classes, immediately related to the two possible
values for orientation. Hereafter we denote   $\vcx = [\ee_1, \ee_2, \ldots, \ee_n]$. 
 The term $\vcx^2$ does not change  sign under orientation change and $\vcx^2 = 1$. 
 The map $\vcx \mapsto -\vcx$ corresponds to a orientation change in $V$ and it is clear that $
 \vcx = [\ee_1, \ee_2, \ldots, \ee_n] = (-1)^{i-1}[\ee_i, \ee_1, \ee_2, \ldots, {\check{\ee_i}},\ldots, \ee_n],$
where ${\check{\ee_i}}$ means that $\ee_i$ is absent from the bracket. Since  there is always a natural correspondence between
a vector space $V$ and its dual $V^*$, all considerations above can be asserted \emph{mutatis mutandis} for $V^*$, 
and hereon  Rota's bracket, now defined as taking values at the dual space, by abuse of notation is also denoted by
$[\;\;]: \underset{n\; \text{times}}{\underbrace{V^*\times V^*\times\cdots\times V^*}} \rightarrow \RR$.  

Consider now a  canonically isomorphic copy of $V^*$,  denoted by $\mrr{V}^*$, with a basis 
$\{\mrr{\ee}^i\}$. This new basis maps vectors in pseudoscalars,
 according to the definition:
\bege
\mrr{\ee}^i(\ee_j) = (-1)^{i-1}[\ee_j, \ee_1, \ee_2, \ldots, {\check{\ee_i}},\ldots, \ee_n] = \vcx\delta^i_j = \vcx \ee^i(\ee_j).\enge
\noi  We can write
\bege\label{queque}
\mrr{\ee}^i = \vcx \ee^i.
\enge \noi Covectors of $\mrr{V}^*$ change sign under orientation change\footnote{An chiral covector $\mrr{\ee}^i$ is defined directly from its action on a vector of $V$ and  \newline
$\mrr{\ee}^i(\,\cdot\,) = (-1)^{i-1}[\;\;\cdot\;, \ee_1, \ee_2, \ldots, {\check{\ee_i}},\ldots, \ee_n]$.}. 
Multiplication by $\vcx$ is clearly an isomorphism between $V^*$ and  $\mrr{V}^*$.

\section{The extended exterior algebra}
\label{sec4} In this section we establish the notion of an extended exterior  algebra, using the pseudoscalar $\vcx$, and also chiral quasi-Hodge star operators
are introduced. 
\subsection{The wedge product from the bracket}

From a Peano (dual) space $V^*$  an exterior algebra can be constructed, by introducing equivalence classes of ordered vector sequences, 
using the bracket \cite{fauser}. Given ${\bf a}^i, {\bf b}^i \in V^*$, two sequences are said to be equivalent, and denoted by  
${\bf a}^1,\ldots, {\bf a}^k \sim  {\bf b}^1,\ldots, {\bf b}^k,$  if for any choice of covectors $\vv^{k+1},\ldots, \vv^n \in V^*$ it follows that  
$[{\bf a}^1,\ldots, {\bf a}^k, \vv^{k+1},\ldots, \vv^n] =  [{\bf b}^1,\ldots, {\bf b}^k, \vv^{k+1},\ldots, \vv^n].$ 
 The wedge\footnote{Grassmann \cite{gra} in his original work called this product the progressive product.} product
between two covectors $\ee^i, \ee^j \in V^*$ is defined as the elements of the quotient space $T(V^*)/J$, where 
$J$ denotes the bilateral ideal generated by elements of the form $a \ot x \ot x \ot b$, where $x \in V$ and $a, b \in T(V)$.
In what follows we write \cite{fauser}
\bege\label{erty}
\ee^i\w\ee^j  = \ee^i\ot \ee^j \;\mod\; \sim
\enge\noi where $\ee^i\ot\ee^j\in V^* \otimes V^*$. For more details see \cite{rota,fauser}. A  
$k$-covector is defined inductively by the wedge product of $k$ covectors, and 
each $k$-covector  lives in  $\Lambda^k(V)$. The exterior algebra is naturally defined
as being $\Lambda(V) = \bigoplus_{k=0}^n \Lambda^k (V)$.  
Analogously  chiral $k$-covectors, elements of $\mrr\la^k(V)$, are defined as the wedge product of elements in $V^*$ and an odd number of elements in $\mrr{V}^*$.
 We also define the chiral
 exterior algebra, which elements change sign under orientation change, as
$\mrr{\Lambda}(V) := \bigoplus_{k=0}^n \mrr{\Lambda}^k (V)$. 
 We denote $ \la^0(V) = \RR$ (scalars), $\mrr\la^0(V) = \vcx\RR$ (pseudoscalars), $\la^1(V) = V^*$ and $\mrr\la^1(V) = \mrr{V}^*$.
The extended exterior algebra  is defined as \bege\breve\la(V) = \la(V) \op \mrr\la(V).\enge \noi  Such an algebra is   
 $\ZZ_2\times\ZZ_2$-graded, where the first $\ZZ_2$-grading is related to the differential forms\footnote{Hereon
it will be implicit that when we refer to differential forms, there is considered a manifold $M$ and its associated 
tangent space [cotangent space] $T_xM \simeq V$ [$T_x^*M \simeq V^*$] at a point $x\in M$. Then a differential form is an element 
of a section $\sec \Lambda(T^*M)$ of the cotangent exterior bundle.} 
 chirality and the second one is related to
the subspaces $\Lambda^p(V)$,  where $p$ is either even  or odd. The inclusions
$$
\la^k(V)\w\la^l(V) \hko \la^{k+l}(V),\qquad \mrr\la^k(V)\w\mrr\la^l(V) \hko \la^{k+l}(V),\qquad 
\mrr\la^k(V)\w\la^l(V) \hko \mrr\la^{k+l}(V)$$
\noi hold. A differential form is said to be {\it chiral} if it is multiplied by $\vcx$, and consequently every chiral form changes sign under orientation change. 
Obviously the exterior algebra $\mrr\la(V)$ is chiral by construction.
 From eq.(\ref{queque}), multiplication by the pseudoscalar $\vcx$ gives a natural isomorphism between $\la(V)$ and $\mrr\la(V)$. 
Achirla forms are mapped in chiral forms through multiplication by  $\vcx$ in such a way that $\mrr\la^k(V)= \vcx\la^k(V)$. 
 From the relation $\vcx^2 = 1$, the set $\{1,\vcx\}$ generates the real algebra $\DDD = \RR\oplus\RR$ of hyperbolic (or perplex, or pseudocomplex, or Study) 
numbers \cite{hes,fj,kel}.
 An element of $\DDD$ can be written
 as  $a + b\vcx, a, b \in\RR$. So the extended exterior algebra  
$${\breve\la}(V) = \la(V) \op \mrr\la(V) = \la(V) \op \vcx\la(V)$$  can be written as 
$\breve\la(V) = \DDD\ot\la(V)$. 
 If a $2n$-dimensional dual space is considered 
($n$ achirla forms and  $n$ chiral forms), then only the exterior algebra generated by the  $n$ achirla forms is needed, since the other (chiral) forms can be generated from 
multiplication by $\vcx$. 

Given an arbitrary basis $\{\ee_i\}$ of $V$, a chiral vector space $\mrr{V}$ is defined to be the vector space spanned by vectors
$\mrr{\ee}_i:=\vcx\ee_i$. In this sense, the same formulation is valid both for vector fields \emph{and}  for differential forms.  

\subsection{Dual chiral quasi-Hodge isomorphisms}

Consider a $n$-vector $\Theta = a\ee_1\w\cdots\w\ee_n$  and a  $n$-covector $\Upsilon = a'\ee^1\w\cdots\w\ee^n$, where
 $a$ and $a'$ are scalars. Denoting $\lrcorner$ the (left) contraction, we have the relation\footnote{Hereafter we denote $\tilde{\psi}$ 
the main anti-automorphism of exterior algebras acting on a form $\psi$.}:
\bege
{\widetilde\Upsilon}\lrcorner\Theta = aa'(\ee^n\w\cdots\w\ee^1)\lrcorner(\ee_1\w\cdots\w\ee_n) = aa',
\enge\noi such that
\[
0\neq  {\widetilde\Upsilon}\lrcorner\Theta = \begin{cases} > 0  &\text{if $a > 0$ and $a'>0$,  or $a<0$ and $a'<0$},\\
                                 < 0 & \text{if $a > 0$ and $a'<0$, or $a<0$ and $a'>0$.}\end{cases}
\]
The orientation of  $V$ can be related to the orientation of the dual  $V^*$. Both orientations of $V$ and $V^*$,
which are respectively determined by  $\Theta$ and $\Upsilon$, are said to be compatible, if ${\widetilde\Upsilon}\lrcorner\Theta > 0$. 
Assuming the orientations  of $V$ and $V^*$ to be compatible,
if we choose an orientation for one of these spaces, the orientation of the another one is completely defined.
 In this case  $\Upsilon$ is chosen such that ${\widetilde\Upsilon}\lrcorner\Theta = 1.$

Denoting $\la_k(V)= \la^k(V^*)$ the space of $k$-vectors, the  dual quasi-Hodge star operators are defined as
\beq
\sa:\la_k(V) &\rightarrow& \la^{n-k}(V)\nonumber\\
      \psi_k &\mapsto& \sa \psi_k = {\widetilde{\psi_k}}\lrcorner\Upsilon
\eeq
\noi ($\sa 1 = \Upsilon$) and
\beq
\saa:\la^k(V) &\rightarrow& \la_{n-k}(V)\nonumber\\
      \psi^k &\mapsto& \saa (\psi^k) = {\widetilde{\psi^k}}\lrcorner\Theta
\eeq\noi ($\saa 1 = \Theta$). It follows that $
\sa\,\saa = \saa\,\sa = (-1)^{k(n-k)}1.
$ Analogously we define the dual chiral quasi-Hodge star operators
\beq
\sa_\vcx:\la_k(V) &\rightarrow& \mrr\la^{n-k}(V)\nonumber\\
      \psi_k &\mapsto& \sa_\vcx (\psi_k) = \vcx{\widetilde{\psi_k}}\lrcorner\Upsilon
\eeq
\noi ($\sa_\vcx 1 = \vcx\Upsilon$) and
\beq
\saa_\vcx:\mrr\la^k(V) &\rightarrow& \la_{n-k}(V)\nonumber\\
      \vcx \psi^k &\mapsto& \saa_\vcx (\vcx \psi^k) = \widetilde{\psi^k}\lrcorner\Theta
\eeq\noi ($\saa_\vcx \vcx = \Theta$). Obviously $\sa_\vcx = \vcx\sa$ and $\saa_\vcx = \vcx\saa$.

\section{The extended Grassmann algebra}
\label{seccla}
By considering an isomorphism $V \simeq V^*$  a correlation is defined to be a  linear map\footnote{Indeed it is a non-canonical isomorphism.}
  $\tau : V \rightarrow V^*$, which induces a (bilinear, symmetric, non-degenerate) 
metric $g : V^* \times V^* \rightarrow \RR$ as 
$g(\ee^i, \ee^j) =  \tau^{-1}(\ee^i)(\ee^j) = g^{ik}\ee_k(\ee^j) = g^{ik}\delta_{k}^j = g^{ij}.$
The extended Grassmann algebra is defined to be the extended exterior algebra endowed with the induced metric.
 Taking two copies of  $V^*$ ($V^*$ and $\mrr{V}^*$),  
 a correlation for each one of these copies is defined:\\
{ \begin{tabular}{lc}
 \begin{minipage}{7cm}
\beq
\tau: V &\rightarrow& V^* \nonumber\\
                  \ee_i&\mapsto& \tau(\ee_i) = g_{ij}\ee^j \nonumber
                  \eeq 
 \end{minipage} &

 \begin{minipage}{7cm}
\beq
\mrr\tau: V &\rightarrow& \mrr{V}^* \nonumber\\
                  \ee_i&\mapsto& \mrr\tau(\ee_i) = \mrr{g}_{ij}\mrr\ee^j = \vcx\mrr{g}_{ij}\ee^j\nonumber
                  \eeq
  \end{minipage}
 \end{tabular}}
\medbreak
\medbreak
\noi
and the associated metrics are given by:\\
{ \begin{tabular}{lc}
 \begin{minipage}{7cm}
\beq
g: {V}^* \times V^* &\rightarrow& \RR \nonumber\\
                  (\ee^i, \ee^j) &\mapsto& g(\ee^i, \ee^j) = g^{ij}\nonumber 
                                   \eeq\noi  
 \end{minipage} &
 \begin{minipage}{7cm}
\beq
\mrr{g}: \mrr{V}^* \times \mrr{V}^* &\rightarrow& \RR \nonumber\\
                  (\mrr\ee^i, \mrr\ee^j) &\mapsto& \mrr{g}(\mrr\ee^i, \mrr\ee^j) = \mrr{g}^{ij}  \nonumber
                  \eeq\noi  
 \end{minipage} 
 \end{tabular}}\medbreak\medbreak

The metrics $\overset{\epsilon}{g}:\mrr{V}^*\times V^*\Ra \RR$ and  $\overset{\backepsilon}{g}:V^*\times \mrr{V}^*\Ra \RR$ are defined to be identically  
null, in such a way that
\bege\label{acl3}
\overset{\epsilon}{g}(\vcx \ee^i, \ee^j) = 0 = \overset{\backepsilon}{g}(\ee^i, \vcx \ee^j).
\enge
Otherwise some inconsistencies arise. 

Now, considering $V\simeq \RR^{p,q}$  the operators
$
\Upsilon:\breve\la(V)\Ra\breve\la(V)
$  admit a fundamental representation 
\bege\label{es22}
\rho(\Upsilon) = \bx \Upsilon_1&\Upsilon_2\\ \Upsilon_3&\Upsilon_4\ex,
\enge\noi acting on elements $\binom{\psi}{{\phi}}$, where $\psi\in\la(V)$ and $\phi\in\mrr\la(V)$.
The operators $\Upsilon_d$ ($d=1,\ldots,4$) are elements of $\clp$ defined by the maps
\beq
\up_1:\la(V)\Ra\la(V),\quad \up_2:\mrr\la(V)\Ra\la(V),\n
\up_3:\la(V)\Ra\mrr\la(V),\quad \up_4:\la(V)\Ra\mrr\la(V).
\eeq

Using the periodicity theorem of Clifford algebras\cite{bt}, that asserts $\cl_{p+1,q+1}\simeq \cl_{p,q}\ot\cl_{1,1}\simeq \cl_{p,q}\ot {\mt M}(2,\CC)$,
it is immediate that $\Upsilon\in {\mt M}(2,\RR)\ot\clp\simeq\cl_{1,1}\ot\cl_{p,q}\simeq
\cl_{p+1,q+1}$, and since it is well known that algebraic spinors associated with $\cl_{p+1,q+1}$ define twistors  \cite{abo, kel},
 ideals of  $\breve\la(V)$ are also useful to describe twistors, at least when $\dim V = 1,2,4$, and consequently to investigate their
profound applications in  physical theories \cite{abo,beng,kel,bet,b1,b2,b3,pe1}.
 
We obtain a representation $\rho: \mrr\la(V)\Ra {\rm End}\,\breve\la(V)$ of the  pseudoscalar $\vcx\in\mrr\la^0(V)$, as
\bege\label{ghji}
\rho(\vcx) = \bx 0&1\\1&0\ex\enge\noi since $\vcx$ changes the chirality of differential forms.  Indeed, given $\psi\in\la(V)$ and $\phi\in\mrr\la(V)$,
\bege
\rho(\vcx)\binom{\psi}{\phi} = \bx 0&1\\1&0\ex \binom{\psi}{\phi} = \binom{\phi}{\psi}\in\la(V)\op\mrr\la(V)\simeq\breve\la(V).
\enge\noi

Denoting the unit of $\RR$ by 1 and the unit of $\mrr{\RR}$ by $\mrr{1}$, 
they are  to be represented respectively by  
\bege\label{uni}
\rho(1) = \bx 1&0\\0&0\ex,\quad \rho(\mrr{1}) = \bx 0&0\\0&1\ex,
\enge\noi where each matrix is an element of $\cl_{p+1,q+1}$ with entries in $\cl_{p,q}$. 
The unit associated with $\RR\op\mrr\RR$ (the field over which $V\op \mrr{V}\simeq \DDD\ot V$ is construted) is given by $\breve{1} = 1 + \mrr{1}$, and can be represented by 
$\rho(\breve{1}) =  \bx 1&0\\0&1\ex$. Basis elements of $V^*$ and $\mrr{V}^*$ are respectively represented as\footnote{Here we transit from 
$V$ $[\mrr{V}]$ to $V^*$ [$\mrr{V}^*$], since there is a (non-canonical) isomorphism between $V$ and $V^*$.}
\bege
\rho(\ee^i) = \bx \ee^i&0\\0&0\ex, \quad \rho(\mrr{\ee}^i) = \bx 0&0\\ 0&\ee^i\ex.
\enge\noi 
When the pseudoscalar $\vcx$ is represented as in eq.(\ref{ghji}), some properties defining the extended Clifford algebras are verified.
The Clifford product is defined in $V^*$ by
\beq\label{acl1}
\ee^i\ee^j + \ee^j\ee^i &=& \rho^{-1}\left[\bx \ee^i&0\\0&0\ex\bx \ee^j&0\\0&0\ex + \bx \ee^j&0\\0&0\ex\bx \ee^i&0\\0&0\ex\right]\n
&=& \rho^{-1}\bx \ee^i\ee^j + \ee^j\ee^i&0\\0&0\ex = \rho^{-1}\bx 2g(\ee^i,\ee^j)&0\\0&0\ex\n  
&=& 2g(\ee^i,\ee^j)\; 1,
\eeq and in $\mrr{V}$ by 
\beq\label{acl2}
\mrr\ee^i\mrr\ee^j + \mrr\ee^j\mrr\ee^i &=& \rho^{-1}\left[\bx 0&0\\ 0&\ee^i\ex \bx 0&0\\ 0&\ee^j\ex +  \bx 0&0\\ 0&\ee^j\ex \bx 0&0\\ 0&\ee^i\ex\right]\n
 &=&\rho^{-1}\bx 0&0\\0&\ee^i\ee^j + \ee^j\ee^i\ex = \rho^{-1}\bx 0&0\\0&2g(\ee^i,\ee^j)\ex\n
 &=& 2g(\ee^i,\ee^j)\;\mrr{1},
\eeq\noi where the metrics in $V^*$ and in $\mrr{V}^*$ can be respectively represented by  
\bege
\rho(g) = \bx g&0\\0&0\ex,\quad \rho(\mrr{g}) = \bx 0&0\\0&g\ex.
\enge\noi The representations of $g$ and $\mrr{g}$ are related by
\bege
\rho(g) = \bx g&0\\0&0\ex = \bx 0&1\\1&0\ex  \bx 0&0\\0&g\ex    \bx 0&1\\1&0\ex = \rho(\vcx)\rho(\mrr{g})\rho(\vcx)^{-1},
\enge\noi what shows that, analogously to conformal transformations in $\RR^{p,q}$ \cite{hest3, mak, cra2, klo, tel}, the metrics associated with spaces of 
different chirality are related by adjoint
representations. 
The extended metric $\breve{g} = g + \mrr{g}$ in $V^*\op \mrr{V}^*$ is given by  
\bege\label{acl4}
\rho(\breve{g}) = \bx g&0\\0&g\ex.
\enge

Chiral and achiral Clifford algebras are then introduced, in the context of the periodicity theorem of Clifford algebras,
 from relations (\ref{acl1}, \ref{acl2}) as 
\bege\label{ac1}
\ee^i\ee^j + \ee^j\ee^i = 2g^{ij}\;1,\qquad 
\mrr\ee^i\mrr\ee^j + \mrr\ee^j\mrr\ee^i = 2\mrr{g}^{ij}\;\mrr{1},
\enge\bege\label{ac3}\ee^i\mrr\ee^j + \mrr\ee^j\ee^i =  0,
\enge\noi where this last relation denotes definition given by eq.(\ref{acl3}).
  
The algebra $\DDD\ot{\cl}_{p,q}$ can be shown not to be a Clifford algebra. Indeed, considering
 $\cl_{1,0}\simeq\DDD\simeq \RR\oplus\RR$, then $\DDD\otimes\cl_{1,0} \simeq \RR\oplus\RR\op\RR\op\RR$, which is not a Clifford algebra.
It can be shown that all subalgebras of a Clifford algebra are either Clifford algebras, or algebras of type  $\DDD\ot\cl_{r,s}$, or 
of type $(\DDD\ot\DDD)\ot \cl_{r,s}$ \cite{mj}. It follows the importance of defining and investigating algebras of type $\DDD\ot\clp$, 
 allowing us  completely to classify all subalgebras of a  Clifford algebra. 
Even though  $\DDD\ot \clp$ is not a  Clifford algebra, as Clifford algebras were also defined in the context given by equations  eqs.(\ref{acl1}, \ref{acl2}), 
it is possible to define a hyperbolic Clifford algebra, over the vector space $V\op\mrr{V}\simeq\DDD\ot V$,  
in the light of the formalism presented in Sec. \ref{secdal}. For more details, see, e.g., \cite{vaz}.

\subsection{(Chiral)  Hodge star operators} 
\label{subsec3.1}

The vector spaces $\la^k(V)$ [$\mrr\la^k(V)$] and $\la^{n-k}(V)$ [$\mrr\la^{n-k}(V)$] have the same dimension, but it does not exist any  canonical isomorphism  
between these spaces. Let $\Theta$ be the volume element in $V$ defined by  $\Theta = |{\rm det} \;\tau|^{1/2} {\bf e}^1 \w \cdots \w {\bf e}^n$, 
where ${\rm det} \;\tau$ is given\footnote{The notation $\tau$ is used to describe the map $\tau : V \rightarrow V^*$ and its natural extension 
 $\tau: \la_k(V) \rightarrow \la^{k}(V)$,\newline without any distinction. It will be implicit which is each one of them in the text.}  implicitly by 
\bege\label{detem}
\tau({\bf e}_1 \w {\bf e}_2 \w \cdots \w {\bf e}_n) = \tau({\bf e}_1) \w \tau({\bf e}_2) \w \cdots \w \tau({\bf e}_n) = ({\rm det} \;\tau)\; {\bf e}_1 \w {\bf e}_2 \w \cdots \w {\bf e}_n.
\enge

The isomorphism given by the dual Hodge star operator  $\star : \la^k(V) \rightarrow \la^{n-k}(V)$ [$\star : \mrr\la^k(V) \rightarrow \mrr\la^{n-k}(V)$] 
is defined from the quasi-Hodge star operators. Since a correlation is defined as an isomorphism  $\tau:\la_k(V)\rightarrow\la^k(V)$ [$\tau:\mrr\la_k(V)\rightarrow\mrr
\la^k(V)$], 
it follows that  
$\sa\circ\tau^{-1}: \la^k(V)\Ra\la^{n-k}(V)$ [$\sa\circ\tau^{-1}: \mrr\la^k(V)\Ra\mrr\la^{n-k}(V)$] \; and \; $\saa\circ\tau: \la_k(V)\Ra\la_{n-k}(V)$
[$\saa\circ\tau: \mrr\la_k(V)\Ra\mrr\la_{n-k}(V)].$  Demand that 
\bege
\sa\circ\tau^{-1} = \saa\circ\tau,
\enge
\noi which occurs only when  $\Theta$ is unitary. The Hodge star operator is defined as
\bege 
\star = \sa\circ\tau^{-1} = \saa\circ\tau
\enge\noi and more explicitly,
\bege\label{stareta}
\star 1 = \eta, \qquad \star \psi = \tau^{-1}(\widetilde{\psi}) \lrcorner \eta,\enge 
\noi where $\psi\in\bla(V)$. The dual Hodge star operator $\star$ does not change the chirality of forms. 
 We define the chiral Hodge star operator as $\star_\vcx : \la^k(V) \rightarrow \mrr\la^{n-k}(V)$  [$\star_\vcx : \mrr\la^k(V) \rightarrow \la^{n-k}(V)]$, given by 
\noi \bege
\star_\vcx 1 = \vcx\eta,\qquad \star_\vcx \psi = \vcx\tau^{-1}(\widetilde{\psi}) \rf \eta.\enge 
\noi We observe that $\star_\vcx = \vcx\star$ and $\star_\vcx$ naturally changes the chirality of forms.

\section{Subspaces embedding and Witt bases}
\label{secdal}
Consider the metric vector space $(V\op \mrr{V}, \ul{g})$,  and denote $u = \uu + \mrr{\uu},\; v = \vv + \mrr{\vv}\in V\op\mrr{V}$.
The metric $\ul{g}:(V\op \mrr{V})\times (V\op\mrr{V})\Ra \mrr\RR$ is given by  
\beq
\ul{g}(u,v) &=& g(\mrr{\uu},\vv) + g(\mrr{\vv},\uu).
\eeq\noi Using the notation introduced in  Sec. \ref{seccla}, the metric $\ul{g}$ can be represented as
$
\bx 0&g\\g&0\ex$. 
 
Under the inclusion maps 

\begin{tabular}{lc}
\begin{minipage}{5cm}
\beq\label{inclu}
i_V : V &\Ra& V\op\mrr{V}\nonumber\\
\vv &\mapsto& \vv + 0\nonumber\eeq\noi\end{minipage} &
\begin{minipage}{5cm} \beq\label{inclu1}
i_{\mrr{V}} :\mrr{V} &\Ra& V \op \mrr{V}\nonumber\\
\mrr{\vv} &\mapsto& 0 + \mrr{\vv}\nonumber,\eeq
\end{minipage}
\end{tabular}\medbreak\noi since  $V$ and $\mrr{V}$ are vector subspaces of $V\op\mrr{V}$, 
it follows that
\bege
\ul{g}(\uu + 0, 0+\mrr\vv) = g(\uu,\mrr\vv),\quad\ul{g}(0+\mrr\uu,\vv+0) = g(\mrr\uu,\vv),\quad \ul{g}(\uu+0,\vv+0) =\ul{g}(0+\mrr\uu,0+\mrr\vv) = 0,\nonumber
\enge\noi 
and then $V$ and $\mrr{V}$ are maximal totally isotropic subspaces of $V\op\mrr{V}$. 
There exists a basis $\{\ee_i\}_{i=1}^n$ of $V$ 
and a basis $\{\mrr\ee_j\}_{j=1}^n$ of $\mrr{V}$,
satisfying  
\bege
\ul{g}(\ee_i, \mrr\ee_j) = \delta_{ij},\quad \ul{g}(\ee_i,\ee_j) = \ul{g}(\mrr\ee_i,\mrr\ee_j) = 0.
\enge\noi Motivated by the results in \cite{vaz}, asserting that
\bege\label{sqrt}
\xi_i = (\mrr\ee_i+\ee_i)/\sqrt{2},\qquad \xi_{i+n} = (\mrr\ee_i - \ee_i)/\sqrt{2},
\enge it is easy to see that the vectors $\{\xi_k\}_{k=1}^{2n}$ span $\mrr\RR^{n,n}$, since for $i,j = 1,\ldots,n$ the relations
\bege
\ul{g}(\xi_i,\xi_j) = -\ul{g}(\xi_{i+n},\xi_{j+n}) = \delta_{ij}.
\enge\noi hold. It is worthwhile to emphasize that $\ul{g}(\xi_i,\xi_{k+n}) = 0, \;\; 1\leq i,j\leq n$.

\section{The regressive product}
\label{secund}
Given a representation of a  $k$-covector $\psi = \aaa^1\w\cdots\w\aaa^k$, and  $\{h_1, h_2, \ldots, h_r\}$ a set of non-negative integers 
such that $h_1 + h_2 + \cdots + h_r = k$, a  {\it split of class  $(h_1, h_2, \ldots, h_r)$ of $\psi$} is defined  as a set of
multicovectors $\{\psi^1,\ldots,\psi^r\}$ such that \cite{rota}
\benu
\item $\psi^i = 1$ if $h_i = 0$ and $\psi^i = \aaa^{i_1}\w\cdots\w\aaa^{i_{h_i}}$, \,$i_1 < \cdots < i_{h_i}$, if $h_i\neq0$;
\item $\psi^i\w \psi^j \neq 0$;
\item $\psi^1\w \psi^2\w\cdots\w \psi^r = \pm \psi$.
\enu
\noi  $(\psi)$ denotes the finite set of all the possible splits of the  $k$-covector $\psi$. The regressive product 
\beq
\vee: \la^k(V)\times \la^l(V) &\rightarrow& \la^{k+l-n}(V)\nonumber\\
                  (\psi^k, \phi^l) &\mapsto& \psi^k \vee \phi^l 
                  \eeq
\noi is defined as \cite{rota} 
\bege\label{regre} \psi^k\vee \phi^l = \sum_{(\psi)}\; [\psi^k_{(1)},\phi^l] \; \psi^k_{(2)} = \sum_{(\phi)}\: [\psi^k,\phi^l_{(2)}]\;\phi^l_{(1)}, \qquad {\rm if} \;\;  k + l \geq n. \enge
\noi When $k+l < n$ we have the trivial case $\psi^k\vee \phi^l = 0$. The bracket calculated between two $k$-coforms 
is defined to be \cite{rota}
\bege[\aaa^1\w\cdots\w \aaa^k,  \bb^1\w\cdots\w \bb^k] = [\aaa^1,\ldots, \aaa^k, \bb^1,\ldots, \bb^k], \;\;{\rm if}
\;\; k + l = n,\enge which is identically  null when $k + l \neq n$. Given  $\psi, \phi, \zeta \in \bla(V)$ the following properties are immediately verified:
\benu
\item $ (\psi\vee \phi)\vee \zeta = \psi\vee (\phi\vee \zeta)$,

\item $\psi^{[k]}\vee \phi^{[l]} = (-1)^{[k][l]}\phi^{[l]}\vee \psi^{[k]}, \qquad [i]:= n-i$,

\item $(\psi+\phi)\vee \zeta = \psi\vee \zeta + \phi\vee \zeta,\qquad  \psi\vee(\phi+\zeta) = \psi\vee \phi + \psi\vee \zeta$,

\item $\psi\vee(a\phi) = (a\psi)\vee \phi = a(\psi\vee \phi), \qquad a\in\RR.$ \enu

 The following relations
\bege \ee^i\vee (\ee^1\w\cdots\w \ee^n) = [1, \ee^1\w\cdots\w \ee^n] \ee^i = [\ee^1,\ldots,\ee^n]\ee^i = \vcx\ee^i,\enge
\noi 
\beq (\ee^i\w\ee^j) \vee (\ee^1\w\cdots\w \ee^n) &=& [1, \ee^1\w\cdots\w \ee^n] (\ee^i\w\ee^j)\nonumber\\
                                                 &=& [\ee^1,\ldots,\ee^n](\ee^i\w\ee^j)=  \vcx(\ee^i\w\ee^j),\eeq

\noi and
\beq
(\ee^{i_1}\w\ee^{i_2}\w\cdots\w\ee^{i_k})\vee (\ee^1\w\cdots\w\ee^n) &=& [1,\ee^1\w\cdots\w\ee^n]\,\ee^{i_1}\w\ee^{i_2}\w\cdots\w\ee^{i_k}\nonumber\\
&=&[\ee^1,\ldots,\ee^n]\,\ee^{i_1}\w\ee^{i_2}\w\cdots\w\ee^{i_k}\nonumber\\
&=&\vcx \ee^{i_1}\w\ee^{i_2}\w\cdots\w\ee^{i_k}.
\eeq
\noi hold. From the above expressions it can be proved, proceeding by induction, and from the  linearity of $\psi\in \bla(V)$, that:
\bege
\psi\vee(\ee^1\w\cdots\w \ee^n) = \vcx \psi.
\enge
\noi Using the relation
$$ \ee^i\vee (\ee^1\w\cdots\w{\check{\ee^j}}\w\cdots\w \ee^n) = 
[\ee^i, \ee^1,\ldots,{\check{\ee^j}},\ldots, \ee^n] = \delta^{ij} (-1)^{i-1}\vcx,$$ \noi when $i = j$, the pseudoscalar is represented as
 $\vcx$ by
\bege
\vcx = (-1)^{i-1}\;\ee^i\vee (\ee^1\w\cdots\w{\check{\ee^i}}\w\cdots\w \ee^n).
\enge

\section{Differential coforms and the chiral counterspace}
From the definition of the regressive product, it is immediate that 
\bege\label{nana}
\la^{n-[k]}(V) \vee \la^{n-[l]}(V) \hko \la^{n-([k]+[l])}(V),
\enge
\noi $[i]:=n-i$. The $k$-counterspace $\bigvee^k$ is defined \cite{con} as being
\bege\label{veee}
\bigvee^k = \la^{n-k}(V)
\enge
\noi and under the regressive product it can define a (regressive) exterior algebra. Indeed, it follows from eq.(\ref{nana}) that
$
\bigvee^r\vee\bigvee^s\hko  \bigvee^{r+s}.
$ The  coexterior algebra is then defined as \bege
\bigvee = \bigvee^0\oplus \bigvee^1\oplus\cdots\oplus \bigvee^n = \bigoplus_{j=0}^n \bigvee^j\enge\noi 
which is an exterior algebra with respect to the regressive product. From definition given by eq.(\ref{veee})
 we see that  $\bigvee^1 = \la^{n-1}(V)$, and  ($n-1$)-forms can be seen as 
1-forms associated with  the counterspace $\bigvee^1$. A basis for $\bigvee^1$, denominated cobasis, is defined as the set  
$\{\mme^i\}$ whose elements are defined as \cite{rota,brow}
\bege
\mme^i = (-1)^{i-1}\ee^1\w\cdots\w{\check\ee^i}\w\cdots\w\ee^n
\enge \noi Elements of $\bigvee^1$  are called 1-coforms.  Rota \cite{rota} denominated the algebra  ($\la(V),\w,\vee$)
by  dialgebra or double algebra. This concept is extended by considering the algebra ($\bla(V), \w, \vee$). From the
 definition above it can be seen that $\ee^i\w\mme^i = \ee^1\w\ee^2\w\cdots\w\ee^n.$
Analogously to  eq.(\ref{queque}), a chiral cobasis $\{\mrr\mme^i\}$, whose 
elements  are defined by the identity $\mrr\mme^i = \vcx\mme^i$, can be introduced. 
The unit of the associative algebra generated by  1-coforms is the volume element 
$\ee^1\w\ee^2\w\cdots\w\ee^n$. 
The following proposition is a straightforward generalization of the Grassmann-Rota one, yielding information about the chirality of differential forms and coforms. 
\medbreak
 {\bf Proposition 1}: $\vvn$ $\mme^1\vee\mme^2\vee\cdots\vee\mme^i = \vcx^{i+1}\ee^{i+1}\w\cdots\w\ee^n$ $\vbn$. 
\medbreak
\noi {\bf Proof}: Let \;$\mme^1 = \ee^2\w\ee^3\w\cdots\w\ee^n$ and  $\mme^2 = -\ee^1\w\ee^3\w\cdots\w\ee^n$ be two 1-coforms. It follows that
\beq
\mme^1\vee\mme^2 &=& -(\ee^2\w\ee^3\w\cdots\w\ee^n)\vee(\ee^1\w\ee^3\w\cdots\w\ee^n)\nonumber\\
                 &=& -[\ee^2,\ee^1,\ee^3,\ldots,\ee^n]\ee^3\w\cdots\w\ee^n\nonumber\\
                 &=& \vcx \ee^3\w\cdots\w\ee^n.\nonumber
\eeq 

\noi We now proceed by induction:
\beq
\mme^1\vee\mme^2\vee\cdots\vee\mme^i\vee\mme^{i+1} &=& (\vcx^{i+1}\ee^{i+1}\w\cdots\w\ee^n)\vee\mme^{i+1}\nonumber\\ &=& 
\vcx^{i+1}(\ee^{i+1}\w\cdots\w\ee^n)\vee((-1)^{i}\ee^1\w\cdots\w{\check{\ee^{i+1}}}\w\cdots\w\ee^n)\nonumber\\
&=& (-1)^i\vcx^{i+1}[\ee^{i+1},\ee^1\w\cdots\w{\check{\ee^i}}\w\cdots\w\ee^n]\,\ee^{i+2}\w\cdots\w\ee^n\nonumber\\
&=& \vcx^{i+1}[\ee^{i+1},\ee^1,\ldots,{\check{\ee^i}},\ldots,\ee^n]\,\ee^{i+2}\w\cdots\w\ee^n\nonumber\\
&=&\vcx^{i+1}\vcx\ee^{i+2}\w\cdots\w\ee^n\nonumber\\
&=&\vcx^{i+2}\ee^{i+2}\w\cdots\w\ee^n\qquad\qquad\qquad\qquad\qquad\square\nonumber\eeq
\medbreak
\noi Depending on the number ($i$) of elements of the product  $\mme^1\vee\mme^2\vee\cdots\vee\mme^i$,
the RHS of Prop. 1 changes or does not change sign under orientation change. As a corollary, when $i=n$, we obtain
\bege
\mme^1\vee\mme^2\vee\cdots\vee\mme^n = \vcx^{n+1}\in \begin{cases} &\text{$\la^0(V)$, if $n$ = 2$k$ + 1,}\\
                                &\text{$ \mrr\la^0(V)$, if $n$ = 2$k$.}\end{cases}\enge

\noi So the volume element  (under the regressive product) $\mme^1\vee\mme^2\vee\cdots\vee\mme^n \in \bigvee^n$ is a scalar or a pseudoscalar,
depending on the dimension  $n$ of $V$. We conclude from these considerations that
\bege\label{ritaguedes}
\bigvee^0 \oplus \bigvee^1\oplus\bigvee^2\oplus\cdots\oplus\bigvee^n = \la^n(V) \oplus \mrr\la^{n-1}(V)\oplus\mrr\la^{n-2}\oplus\cdots\oplus
\la^0 [\mrr\la^0](V), 
\enge
\noi where the last term of the direct sum above denotes the two possibilities, depending on the even or odd value of $n$. 
From Prop. 1 and the properties of the dual Hodge star operator, the relation
\bege
\star (\ee^1\w\ee^2\w\cdots\w\ee^k) =  \vcx^{k+1}\mme^1\vee\cdots\vee\mme^k
\enge
\noi follows. The Hodge star operator applied on ($2k$)-forms gives chiral ($2k$)-coforms, and when it is applied on ($2k$+1)-forms, its
 respective ($2k$+1)-coforms have no chirality ($2k$ + 1 $\leq n$).

\section{Clifford algebras over the counterspace}

In this section the formulation given by Conradt \cite{con} is reinterpreted, and his definitions are slightly modified in order to encompass 
the present formalism.
It is well known that given a  volume element $\eta\in\Lambda^n(V)$, the dual Hodge star operator  
acting on a multivector $\psi$ can be defined as $\star\psi =  \tilde\psi \eta$ and $\star 1 = \eta$. 
The Clifford product $\ast: \clp\times\clp \Ra \clp$,  related to the counterspace is defined as:
\bege\label{newp}
{}{\psi\ast\phi:= \star^{-1}[(\star\psi)(\star\phi)]} \qquad \psi,\phi\in\clp.
\enge\noi Such a product is immediately shown to satisfy a Clifford algebra. First of all the associativity is verified \cite{con}.
Indeed, given  
$\psi,\phi,\zeta\in\clp$, we have:
\beq
(\psi\ast\phi)\ast \zeta &=& \{\star^{-1}[(\star\psi)(\star\phi)]\}\ast \zeta\n
&=& \star^{-1}\{\star\star^{-1}[(\star\psi)(\star\phi)](\star \zeta)\}\n
&=& \star^{-1}\{[(\star\psi)(\star\phi)](\star \zeta)\}\n
&=& \star^{-1}\{(\star\psi)[(\star\phi)(\star\zeta)]\}\n
&=& \star^{-1}\{(\star\psi)\star[\star^{-1}((\star\phi)(\star\zeta))]\}\n
&=&\psi\ast [\star^{-1}((\star\phi)(\star\zeta))]\n
&=& \psi\ast(\phi\ast\zeta).\eeq\noi The additive distributivity, 
\beq
\psi\ast(\phi + \zeta) &=& \psi\ast\phi + \psi\ast \zeta\label{esta11}\\
(\psi+\phi)\ast\zeta &=& \psi\ast\zeta + \phi\ast\zeta,\label{esta12}
\eeq\noi can be also verified. Indeed:
\beq 
\psi\ast(\phi + \zeta) &=& \star[(\star\psi)\star(\phi + \zeta)]\n
&=& \star^{-1}[(\star\psi)(\star\phi + \star\zeta)]\n
&=& \star^{-1}[(\star\psi)(\star\phi)]  +  \star^{-1}[(\star\psi)(\star\zeta)]\n
&=& \psi\ast \phi + \psi\ast\zeta.
\eeq\noi Eq.(\ref{esta12}) is shown in an analogous way.

The volume element $\eta$ acts as a unit in relation to the product $\ast$. In fact, $\eta$ is the left unit
\beq
\eta\ast\psi &=& \star^{-1} [(\star\eta)(\star\psi)]\n
&=&   \star^{-1} (1 \star\psi) = \star^{-1}(\star\psi)\n
&=& \psi
\eeq\noi and analogously  $\eta$ is also the right unit related to the product  $\ast$, since 
\beq
\psi\ast\eta &=& \star^{-1} [(\star\psi)(\star\eta)]\n
&=&   \star^{-1} (\star\psi 1) = \star^{-1}(\star\psi)\n
&=& \psi.
\eeq\noi
It is worthwhile to emphasize that both the usual Clifford product, denoted by  juxtaposition, and 
the another Clifford product $\ast: \clp\times\clp \Ra \clp$, related itself to the counterspace, act on the underlying vector space of $\clp$. Since  
the Clifford algebra, constructed from the usual Clifford product is denoted by  $\cl(\Lambda^1(V),g)$, the `another' Clifford algebra is denoted by
 $\cl(\bigvee^1,g)$ when interpreted as being constructed from the $\ast$-product. Indeed, the Clifford relation, computed from the product given by eq.(\ref{newp}) 
between two coforms $\mme_i, \mme_j\in\bigvee^1$ is given by
\beq
\mme^i\ast\mme^j + \mme^j\ast\mme^i &=& \star^{-1}[(\star\mme^i)(\star\mme^j)] + \star^{-1}[(\star\mme^j)(\star\mme^i)]\n
&=&  \star^{-1}(\ee^i\ee^j) + \star^{-1}(\ee^j\ee^i)\n   
&=& \star^{-1}(\ee^i\ee^j + \ee^j\ee^i)\n
&=& \star^{-1}(2g(\ee^i,\ee^j))\n
&=& 2g(\ee^i,\ee^j)\,\eta,
\eeq\noi from eq.(\ref{stareta}). As  $\eta$ is the unit related to the product $\ast$, the product $\ast$ indeed defines a Clifford algebra.

Given $\vv\in V, \psi\in\clp$, the regressive product defined in Sec. \ref{secund} can now be written in terms of the Clifford product $\ast$, as:
\bege
{}{\mme^i\vee\psi = \me(\mme^i\ast\psi + \hat{\psi}\ast\mme^i)}.
\enge\noi The right contraction of a vector $\vv$ by an element  
$\psi\in\clp$, 
associated to the product 
 $\ast$ is defined by \cite{con} 
\bege\label{odor}
{}{\vv\ulcorner\psi := \me(\vv\ast\psi - \hat{\psi}\ast\vv),}
\enge\noi while the left contraction is introduced by the expression
\bege\label{es1}
{}{\psi\urcorner\vv := \me(\psi\ast\vv - \vv\ast\hat{\psi}).}
\enge\noi

\section{Duality and coduality}
Consider two differential forms  ${}\xi\in\Lambda^i(V)$ and $\omega\in\la^j(V)$. The relations 
 $\saa({}\xi\w{}\omega) = (\saa \,{}\xi)\vee (\saa \,{}\omega)$, 
 $\saa({}\xi\vee{}\omega) = (\saa \,{}\xi)\w (\saa\, {}\omega)$, and the same assertions to the operator 
$\sa$ are easily shown \cite{rota}. If we work with a  Grassmann algebra, where a metric is introduced, instead of a Grassmann-Cayley algebra, it is possible to prove
that
\bege\label{drip}
{}{\star({}\xi\w{}\omega) = (\star\,{}\xi)\vee(\star\,{}\omega)}, \qquad
{}{\star({}\xi\vee{}\omega) = (\star\,{}\xi)\w(\star\,{}\omega)}.\enge 
\noi 
Such relations given by eqs.(\ref{drip}), 
besides being shown inside a formalism devoid of indices and/or components, have origin in the definition of  
the Clifford product $\ast$ associated with counterspace. Indeed,
\beq
\xi\vee\omega &=& \langle \xi\ast\omega\rangle_{n-(i+j)}\n
&=&\langle \xi\ast\omega \eta^{-1}\eta\rangle_{n-(i+j)}\n 
&=&\langle \star^{-1}[(\star\xi)(\star\omega)]\eta^{-1}\,\rangle_{i+j}\,\eta\n 
&=& \langle \star^{-1}\{\star[\widetilde{(\star\xi)(\star\omega)}]\}\eta\eta^{-1}\,\rangle_{i+j}\,\eta\n 
&=& \langle \widetilde{(\star\xi)(\star\omega)}\rangle_{i+j}\,\eta\n 
&=&  (\widetilde{\star\omega})\w(\widetilde{\star\xi})\eta\n 
&=&  \widetilde{(\star\xi)\w(\star\omega)}\eta\n 
&=&  \star^{-1}[(\star\xi)\w(\star\omega)] 
\eeq\noi and it follows by linearity that  
\bege\label{dripo}
{}{\star(\xi\vee\omega) =  (\star\xi)\w(\star\omega)}\quad \forall \xi,\omega\in\la(V),
\enge\noi which is exactly  eq.(\ref{drip}). Using the same procedure it is possible to derive eq.(\ref{drip}) as an identity  involving the Clifford product $\ast$. 
Therefore the duality between Clifford algebras over $\bigvee^1$ and $\la^1(V)$
reflects the duality between the spaces  $\bigvee^1$ and $\la^1(V)$. Besides there is also a duality related to the contraction 
 between differential forms and differential coforms, 
where the last was defined by eq.(\ref{odor}). This duality is presented by the following propostion:
\medbreak
 {\bf Proposition 2}: $\vvn {}{\star(\xi\ulcorner\omega) =  (\star\xi)\lrcorner(\star\omega)}\quad  \forall \xi,\omega\in\la(V).\vbn$
\medbreak
\noi {\bf Proof}: 
The definition of the product $\xi\ulcorner\omega$ is valid only when $\omega$ has degree great or equal to $\xi$. Then,
\beq
\xi\ulcorner\omega &=& \langle \xi\ast\omega\rangle_{n-|i-j|}\n
&=&\langle \xi\ast\omega\,(\eta^{-1}\eta)\rangle_{n-|i-j|}\n 
&=&\langle \star^{-1}[(\star\xi)(\star\omega)]\eta^{-1}\,\rangle_{|i-j|}\,\eta\n 
&=& \langle \star^{-1}\{\star[\widetilde{(\star\xi)(\star\omega)}]\}\eta\eta^{-1}\,\rangle_{|i-j|}\,\eta\n 
&=& \langle \widetilde{(\star\xi)(\star\omega)}\rangle_{|i-j|}\,\eta\n 
&=&  (\widetilde{\star\omega})\lrcorner(\widetilde{\star\xi})\eta\n 
&=&  \widetilde{(\star\xi)\lrcorner(\star\omega)}\eta\n 
&=&  \star^{-1}[(\star\xi)\lrcorner(\star\omega)] 
\eeq\noi Therefore $\star(\xi\ulcorner\omega) =  (\star\xi)\lrcorner(\star\omega).\qquad\qquad\qquad\quad\qquad\qquad\qquad\quad\Box$
\medbreak
\noi Analogously it can be asserted the following proposition:
\medbreak
 {\bf Proposition 3}: $\vvn {}{\star(\xi\urcorner\omega) =  (\star\xi)\urcorner(\star\omega)}\quad  \forall \xi,\omega\in\la(V)\vbn$
\medbreak \noi This proposition is valid in the case when $\omega$ has degree less or equal to $\xi$. The proof is analogous to the 
Prop. 2. See \cite{con}.

\section{Differential and codifferential operators}
\label{opd}
In this section the spaces $\Lambda(V)$ must be viewed as $\Lambda(T_x^*M)$.
\subsection{Differential operator}
The differential operator $d: \sec \la^k(T^*M) \rightarrow \sec \la^{k+1}(T^*M)$ acts on a multivector $\psi$ as 
      $\psi \mapsto d\psi = (\partial_{i_j}\psi_{i_1\cdots i_k}) dx^{i_j}\w (dx^{i_1}\w\cdots\w dx^{i_k})
                  $ where $M$ is a manifold which cotangent space $T^*_xM$, at $x\in M$, is isomorphic to  $V^*$.

We have seen that, in order to map an achiral  $k$-form to a chiral  $k$-form, it is needed the  multiplication by   $\vcx$. 
Given ${\psi}\in \sec\la^k(T^*M)$, since $d$ is defined to be  ``$\DDD$-linear'', i.e., $d(\vcx \psi^k) = \vcx d \psi^k$, and that  $d\psi^k \in \sec\la^{k+1}(T^*M)$,
 then $\vcx d\psi^k \in \sec\mrr\la^{k+1}(T^*M)$. It follows that  $d:\sec\mrr\la^k(T^*M) \rightarrow \sec\mrr\la^{k+1}(T^*M)$. Motivated by these considerations
 the exterior derivative is defined as the unique set of operators   
 $d:\sec\la^k(T^*M) \rightarrow \sec\la^{k+1}(T^*M)$ and $d:\sec\mrr\la^k(T^*M) \rightarrow \sec\mrr\la^{k+1}(T^*M)$ that satisfy the following properties:
\benu
\item $d(\zeta + \omega) = d\zeta + d\omega,\quad {\rm  and}\quad   d(c\om) = c\,d\om, \;  \forall\zeta, \om\in\sec\bla(T^*M), c\in\RR,$
\item ${d(\omega\w\zeta) = d\omega\w\zeta + (-1)^k\omega\w d\zeta,\quad  \forall\omega\in\sec\la^k(T^*M\op\mrr T^*M), \zeta\in\sec\bla(T^*M)}\label{antideri}$,
\item $ d(d\om) = 0,\qquad  \forall\omega\in \sec\bla(T^*M).$
\enu\noi By linearity it is possible to extend the definition of $d$ to the extended exterior algebra $d:\sec\bla(T^*M)\rightarrow\sec
\bla(T^*M)$. 

\subsection{Codifferential operator}
Considering $l = n - 1$ in eq.(\ref{nana}), it can be seen that  
\bege \sec\la^k(T^*M) \vee \sec\la^{n -1}(T^*M) \hko \sec\la^{k - 1}(T^*M), \qquad  k \geq 1,
\enge \noi which motivates us to define the codifferential operator from the regressive product as
\beq\label{coddd}
\delta: \sec\la^k(T^*M) &\rightarrow& \sec\la^{k-1}(T^*M) \nonumber\\
                   \psi &\mapsto& \delta \psi = (g_{i_ki_j}\partial^{i_k}\psi_{i_1\cdots i_k}) (dx^{i_1}\w\cdots\w dx^{i_k}) \vee \left[(dx^1\w\cdots\w
{\check{dx^{i_j}}}\w\cdots\w dx^n)\right].\nonumber
                  \eeq\noi From the regressive product associativity it follows that
\bege
{}{\delta(\psi\vee \phi) = \delta \psi \vee \phi + (-1)^{[\psi]} \psi\vee \delta \phi}
\enge\noi where $\psi, \phi\in \bla(T^*M)$, and $[\psi] = k$ if $\psi\in\Lambda^k(T^*M)$ or $\psi\in\sec\mrr\Lambda^k(T^*M)$ .

The counterspace has the codifferential operator  $\delta$ acting as its associated differential operator.
This can be illustrated by the following de Rham sequences:
\bege
\la^0(V)\xrightarrow{d}\la^1(V)\xrightarrow{d}\la^2(V)\xrightarrow{d}\cdots\xrightarrow{d}\la^{n-1}(V)\xrightarrow{d}\la^n(V)\xrightarrow{d} 0,
\enge
\bege
0\xleftarrow{d}\bigvee^0\xleftarrow{d}\bigvee^1\xleftarrow{d}\bigvee^2\xleftarrow{d}\cdots\xleftarrow{d}\bigvee^{n-1}
\xleftarrow{d}
\bigvee^n
\enge 

\bege
\bigvee^0\xrightarrow{\delta}\bigvee^1\xrightarrow{\delta}\bigvee^2\xrightarrow{\delta}\cdots\xrightarrow{\delta}\bigvee^{n-1}
\xrightarrow{\delta}
\bigvee^n\xrightarrow{\delta} 0,
\enge
\bege
0\xleftarrow{d}\la^0(V)\xleftarrow{\delta}\la^1(V)\xleftarrow{\delta}\la^2(V)\xleftarrow{\delta}\cdots\xleftarrow{\delta}
\la^{n-1}(V)\xleftarrow{\delta}\la^n(V).\enge
\noi

\subsection{The Hodge-de Rham Laplacian}

The  Laplacian $\Delta$ is naturally defined as
\bege\label{lap}
{}{\Delta = d\delta + \delta d}
\enge\noi We exhibit a simple example:
\medbreak
{\bf Example}: Consider $\psi\in\sec\la^2(T^*\RR^3)$, given  $\psi = f(x^1,x^2,x^3)dx^1\w dx^2$, where $f$ is a scalar field $f:\RR^3\Ra \RR$. It follows that
\bege
d\psi = \frac{\pa f}{\pa x^3} dx^1\w dx^2\w dx^3.
\enge\noi  Therefore, 
\beq\label{deld}
\delta d\psi &=& \frac{\pa^2 f}{\pa x^1\pa x^3} (dx^1\w dx^2\w dx^3)\vee (dx^2\w dx^3) + \frac{\pa^2 f}{\pa x^2\pa x^3} (dx^1\w dx^2\w dx^3)\vee (dx^1\w dx^3)\n
&& + \frac{\pa^2 f}{\pa (x^3)^2} (dx^1\w dx^2\w dx^3)\vee (dx^1\w dx^1) \n
 &=& \frac{\pa^2 f}{\pa x^1\pa x^3}(dx^2\w dx^3) + \frac{\pa^2 f}{\pa x^2\pa x^3} (dx^3\w dx^1) + \frac{\pa^2 f}{\pa (x^3)^2}  (dx^1\w dx^2) 
\eeq\noi On the other hand, 
\beq
\delta\psi &=&  \frac{\pa f}{\pa x^1} (dx^1\w dx^2)\vee (dx^2\w dx^3) + \frac{\pa f}{\pa x^2} (dx^1\w dx^2)\vee (dx^1\w dx^3) + 
\frac{\pa f}{\pa x^3} (dx^1\w dx^2)\vee (dx^1\w dx^2)\n
&=& \frac{\pa f}{\pa x^1} dx^2 + \frac{\pa f}{\pa x^2} (-dx^1) + 0.
\eeq\noi It follows that
\beq\label{ddel}
d\delta\psi = \frac{\pa^2 f}{\pa (x^1)^2}(dx^1\w dx^2) + \frac{\pa^2 f}{\pa (x^2)^2} (-dx^2\w dx^1) + \frac{\pa^2 f}{\pa x^1\pa x^3}  dx^3\w dx^2 
+  \frac{\pa^2 f}{\pa x^2\pa x^3}  dx^3\w dx^1.
\eeq From eqs.(\ref{deld}, \ref{ddel}) we have:
\beq
(d\delta + \delta d)\psi &=&  \frac{\pa^2 f}{\pa (x^1)^2} +  \frac{\pa^2 f}{\pa (x^2)^2} +  \frac{\pa^2 f}{\pa (x^3)^2}\n
&=& \Delta\psi.
\eeq\noi It can be shown by induction that eq.(\ref{lap}) is valid for all $\psi\in\sec\la(T^*M)$.

\section*{Concluding remarks}
Peano spaces are the natural arena to introduce extended exterior, Grassmann, and subsequently Clifford algebras
in the light of the regressive product. Besides endowing exterior algebras with chirality, Rota's bracket is suitable 
to define extended exterior, Grassmann, and Clifford algebras, naturally presenting a $\ZZ_2\times\ZZ_2$-graded structure.
The introduction of different units providing the construction of respectively achiral and chiral algebras foresees us 
to use the periodicity theorem of Clifford algebras, asserting that $\cl_{p+1,q+1}\simeq\clp\ot\cl_{1,1}$, in order 
to immerge both achiral and chiral Clifford algebras $\clp$ into $\cl_{p+1,q+1}$. It gives rise to  various 
possibilities of applications in physical theories, like e.g. twistor theory and conformal field theory. In such embedding, 
the extended Clifford algebra associated with $\clp$ is shown to be $\cl_{p+1,q+1}$, wherein the formulation is 
more simple and natural. Moreover, Proposition 1 describes a chiral relationship between differential coforms under the regressive product and forms under the progressive
product. When the regressive product is used, the dual Hodge star operator acts on $k$-forms, resulting
 in a $k$-coform intrinsically endowed with chirality only if $k$ is an even integer, otherwise its action results in an achiral form.  Moreover, the counterspace volume
element with respect to the regressive product is scalar or pseudoscalar until we specify whether the dimension of the Peano space
is respectively odd or even. The $\ast$-Clifford product \cite{con} completes the dual characterization of the counterspace. 
We also introduced pseudoduality between space and counterspace, since 
 the de Rham cochain, generated by the codifferential operator related to the regressive product,  is composed  
by a  sequence of exterior algebra homogeneous subspaces that are subsequently chiral and achiral. This is an astonishing  
character of the formalism to be presented, since the duality between exterior algebras associated respectively with  
the space and counterspace is irregular,
in the sense that if we take the exterior algebra duality associated with the space, we obtain the exterior algebra associated with the counterspace, but 
the converse produces the space exterior algebra, which homogeneous even [odd] subspaces  are chiral [achiral], depending on the original vector space
dimension (see eq.(\ref{ritaguedes})). Then, duality between space and counterspace is deduced to be a pseudoduality 
if the exterior algebra is endowed with chirality.

\end{document}